\documentclass[11pt]{article}
\usepackage[utf8]{inputenc}
\usepackage[T1]{fontenc}
\usepackage{amsmath,amssymb,amsthm}
\usepackage{geometry}
\usepackage{mathtools}
\usepackage{enumitem}
\usepackage{xcolor}
\usepackage{hyperref}
\newtheorem{remark}{Remark}
\hypersetup{colorlinks=true,linkcolor=blue!45!black,citecolor=blue!45!black,urlcolor=blue!45!black}

\newcommand{\dd}{\mathrm{d}}
\newcommand{\half}{\tfrac{1}{2}}
\newcommand{\Aux}{A_{\mu\nu}}
\DeclareMathOperator{\Tr}{Tr}

\theoremstyle{plain}

\title{\textbf{The auxiliary-metric formulation of
Born--Infeld New Massive Gravity}}
\author{Bayram Tekin\footnote{bayram.tekin@bilkent.edu.tr}\\
\small Department of Physics Bilkent University, 06800 Ankara, T\"urkiye}
\date{\today}

\begin{document}
\maketitle

\begin{abstract}
Born--Infeld New Massive Gravity (BINMG) completes New Massive Gravity to all
orders in curvature through the determinant of the metric shifted by the Einstein
tensor. We recast it with an independent auxiliary metric $q_{\mu\nu}$, whose
algebraic equation of motion $q_{\mu\nu}=g_{\mu\nu}+\frac{\sigma}{m^2}
G_{\mu\nu}(g)$ recovers the determinant action exactly on the regular branch and
resums the infinite curvature series into a single relation. In the densitized
variable $P^{\mu\nu}=\sqrt{-q}\,q^{\mu\nu}$ the three-dimensional action is
polynomial, with all derivative dependence carried by the coupling
$P^{\mu\nu}G_{\mu\nu}(g)$. The formulation makes known properties follow with substantially less algebra: the
unique vacuum follows in one line, and the quadratic action yields a single
Pauli--Fierz massive spin-2 field with the Fierz--Pauli tuning generated rather
than imposed. On locally AdS backgrounds, the conserved charges, BTZ mass and
angular momentum, central charge, and entropy reduce to the Einstein results
times a common factor. The formulation also isolates the nonlinear
degree-of-freedom problem in the right variables, leaving the full Dirac count to
separate work.
\end{abstract}

\tableofcontents

\section{Introduction}

Three-dimensional gravity occupies a special place in the study of gravitational
dynamics. The Einstein--Hilbert action in three dimensions admits black-hole
solutions when the cosmological constant is negative~\cite{BTZ}, and is
equivalent to a Chern--Simons theory of a non-compact gauge group~\cite{Witten},
but it propagates no local graviton: the Weyl tensor vanishes identically, and the
Riemann tensor is fixed algebraically by the Ricci tensor, so the vacuum equations force locally constant curvature. Local dynamics must therefore be
supplied by a deformation of the Einstein theory.

Massive gravity provides such a deformation \cite{Deser1}. Topologically massive gravity adds a
parity-violating gravitational Chern--Simons term and yields a single massive
helicity mode~\cite{DJT}, with characteristic parity-odd
features~\cite{causality}. A parity-even alternative was given by the New Massive
Gravity (NMG) of Bergshoeff, Hohm, and Townsend~\cite{BHT1,BHT2}, which adds the
particular curvature-squared invariant
\begin{equation}
  K := R_{\mu\nu}R^{\mu\nu} - \tfrac{3}{8}R^2 ,
  \label{eq:K}
\end{equation}
chosen so that the linearized theory is the Pauli--Fierz theory of a massive
spin-two field. In three dimensions, such a field carries two local degrees of
freedom. The special trace structure of the field equations coming from~\eqref{eq:K} is what removes the
would-be scalar that a generic curvature-squared action would propagate.

Born--Infeld New Massive Gravity (BINMG)~\cite{BINMG} replaces the finite
higher-curvature deformation by the determinant of a curvature-deformed metric,
$g_{\mu\nu}+\tfrac{\sigma}{m^2}G_{\mu\nu}$, where $G_{\mu\nu}$ is the Einstein
tensor. Its small-curvature expansion reproduces NMG at quadratic order, and at
cubic  and higher orders, it reproduces precisely the deformation of NMG obtained from
AdS/CFT~\cite{BINMG,Sinha,cfunctions,Paulos1,Alkac}. It also appeared as a boundary term in four dimensions \cite{Sinha2}. The
Born--Infeld structure packages this infinite tower compactly and preserves the
massive spin-two spectrum about its unique maximally symmetric vacuum. The unitarity
of the metric fluctuations, the conserved charges, and the $c$-functions of
BINMG have been established previously by direct and often lengthy computations
in the determinant form~\cite{BINMG,cfunctions,unitarity,BImassless4d,BIunique,Tahsintez}: the uniqueness of the
vacuum, the absence of a ghost in the linearized spectrum, and the value of the
central charge are known results. Indirect evidence that BINMG belongs to a
ghost-free class of three-dimensional higher-curvature gravities comes from
scaling limits of ghost-free bigravity~\cite{PaulosTolley,deRham3d}.

The contribution of the present paper is not to reestablish these facts but to
provide a \emph{another vantage point} from which they become transparent and from
which further computations follow with little effort. What was previously
obtained through extended curvature expansion, we recover 
and organize through a single structural device: an auxiliary metric.

The determinant form of the theory is elegant; however, it is not the most convenient form for explicit
computation. The deformation tensor sits inside a square root, and its expansion
contains arbitrarily high powers of the curvature. Quantities that are routine in
Einstein gravity --- for example conserved charges, central charges, and black-hole
entropy --- become laborious when computed from the determinant directly, and the
canonical structure is obscured.

In this paper, we develop the \emph{auxiliary-metric formulation} of BINMG and
show that it removes these obstructions. The idea is useful for Born-Infeld
type actions: an independent auxiliary metric linearizes the square root, with an
algebraic equation of motion that returns the original determinant on shell. What
is specific to BINMG is that the auxiliary equation takes the clean form
\begin{equation}
  q_{\mu\nu} = g_{\mu\nu} + \frac{\sigma}{m^2}G_{\mu\nu}(g),
\end{equation}
so that a single algebraic relation resums the entire higher-curvature series.
The resulting action is a second-order two-metric system, structurally close to
bimetric gravity but with one metric algebraically tied to the other. The equivalence between the determinant and auxiliary formulations is understood
on the regular Born--Infeld branch. By this, we mean the sector of field space
in which the curvature-deformed tensor
\[
A_{\mu\nu}(g)
:=
g_{\mu\nu}
+
\frac{\sigma}{m^2}G_{\mu\nu}(g)
\]
is nondegenerate and has Lorentzian signature. In this sector,
\(A_{\mu\nu}\) possesses a well-defined inverse, its determinant has the sign
required for the Born--Infeld square root to be real, and the algebraic
equation
\[
q_{\mu\nu}=A_{\mu\nu}(g)
\]
defines a nondegenerate Lorentzian auxiliary metric. The regularity condition
is therefore not an additional dynamical equation; it specifies the domain on
which the auxiliary description is well defined and algebraically equivalent
to the original determinant theory. Configurations for which
\(\det A_{\mu\nu}=0\), or for which the signature of \(A_{\mu\nu}\) changes,
belong to singular branches and are not covered by the equivalence established
here. The special coupling \(\beta=0\), for which the maximally symmetric
auxiliary metric degenerates, is discussed separately in Section~5 and is
excluded from the regular massive branch.

We use this formulation for several purposes. First, we make the unique maximally
symmetric vacuum of BINMG manifest and contrast it with the two degenerate vacua
of NMG. Second, we compute the linearized spectrum directly from the auxiliary
action, recovering one Pauli--Fierz massive spin-two field; notably, the
Fierz--Pauli mass tuning emerges as an output of the Born--Infeld potential
rather than being imposed. Third, we introduce the densitized inverse auxiliary
metric $P^{\mu\nu}:=\sqrt{-q}\,q^{\mu\nu}$, for which the three-dimensional action
becomes polynomial. Fourth, we show in one representative application that conserved charges,
BTZ thermodynamics, and the Brown--Henneaux central charge reduce on locally AdS
backgrounds to rescaled Einstein computations.  Other applications, such as
boundary terms and holographic $c$-functions, are best developed separately.

The nonlinear Hamiltonian degree-of-freedom count is the natural next application
of this formulation, and indeed, the auxiliary form was constructed with that
problem in mind. That analysis is involved enough to warrant separate treatment,
and we defer it \cite{BTekin}; in Section~\ref{sec:nonlinear-outlook} we state precisely how the
auxiliary form reduces the problem and what remains to be done. The present paper
establishes the formulation and demonstrates its utility on problems where the
answer can be obtained cleanly and checked.

\section{New Massive Gravity and its auxiliary field}

We first recall the NMG auxiliary construction, both for completeness and because the comparison with BINMG is instructive. NMG is
defined by \cite{BHT1,BHT2}
\begin{equation}
  I_{\rm NMG}[g] = \frac{1}{\kappa^2}\int \dd^3x\,\sqrt{-g}\,
  \left[\sigma R - 2\lambda_0 m^2 + \frac{1}{m^2}\Big(R_{\mu\nu}R^{\mu\nu}
  - \tfrac{3}{8}R^2\Big)\right],
  \label{eq:NMG}
\end{equation}
where $\sigma = \pm 1$ and $m^2 >0$.
The fourth-order metric equations conceal a second-order massive spin-two
structure, which an auxiliary symmetric tensor $f_{\mu\nu}$ makes manifest:
\begin{equation}
  I_{\rm NMG}[g,f] = \frac{1}{\kappa^2}\int \dd^3x\,\sqrt{-g}\,
  \left[\sigma R + f^{\mu\nu}G_{\mu\nu}[g]
  - \frac{m^2}{4}\big(f_{\mu\nu}f^{\mu\nu} - f^2\big)\right].
  \label{eq:NMGaux}
\end{equation}
The field $f_{\mu\nu}$ enters without derivatives; its algebraic equation of
motion is
\begin{equation}
  f_{\mu\nu} = \frac{2}{m^2}\Big(R_{\mu\nu} - \tfrac14 g_{\mu\nu}R\Big),
\end{equation}
and substituting it back reproduces~\eqref{eq:NMG}. Two features are essential.
The coupling $f^{\mu\nu}G_{\mu\nu}$ lowers the derivative order, and the mass term
$-\tfrac{m^2}{4}(f_{\mu\nu}f^{\mu\nu}-f^2)$ is exactly of Fierz--Pauli form, with
an invertible coefficient that gives $f_{\mu\nu}$ a healthy massive spin-two
propagation while keeping the trace non-dynamical. The auxiliary field is, in
effect, the massive graviton. Its equation of motion is a \emph{finite-order}
algebraic relation; NMG is already polynomial in the curvature. Canonical analysis \cite{HDcanonical} of the theory directly in the metric formulation confirms these results. 

\section{Born--Infeld New Massive Gravity}

\subsection{The determinant action}

BINMG is defined by~\cite{BINMG}
\begin{equation}
  I_{\rm BINMG}[g] = -\frac{4m^2}{\kappa^2}\int \dd^3x\,
  \left[\sqrt{-\det\Big(g_{\mu\nu}+\tfrac{\sigma}{m^2}G_{\mu\nu}\Big)}
  - \beta\sqrt{-g}\right],
  \qquad \beta = 1 - \frac{\lambda_0}{2}.
  \label{eq:BINMG}
\end{equation}
What motivated this theory, along with several interesting results about it, can be found in the PhD thesis \cite{Tahsintez}.
It is convenient to introduce  the shifted metric (that appeared in the first section above)
\begin{equation}
  \Aux(g) := g_{\mu\nu} + \frac{\sigma}{m^2}G_{\mu\nu}(g).
  \label{eq:Adef}
\end{equation}
The regular Born–Infeld branch is the branch on which $\Aux$ is nondegenerate
and Lorentzian, so the square root in \eqref{eq:BINMG} is real. The special value
$\beta=0$ (equivalently $\lambda_0=2$) is degenerate and excluded; we discuss it
in Section~\ref{sec:vacuum}. Two features distinguish the deformation tensor from
a naive $g_{\mu\nu}+aR_{\mu\nu}$: it contains the Einstein tensor, whose trace
reversal is what makes the quadratic expansion reproduce the NMG
combination~\eqref{eq:K} rather than a generic curvature square, and it generates
infinitely many powers of the curvature. The field equation of this theory is
cumbersome~\cite{cfunctions}, and one of our goals here is to simplify it. For
comparison with what appears below, let us quote the result.
\begin{equation}
  I_{\rm BINMG} = -\frac{4m^2}{\kappa^2}\int \dd^3x\,\sqrt{-g}\,
  F(R,K,S),
  \label{eq:Faction}
\end{equation}
where
\begin{equation}
  F(R,K,S):=
  \sqrt{1-\frac{\sigma}{2m^2}\!\left(R+\frac{\sigma}{m^2}K
  -\frac{1}{12m^4}S\right)}-\left(1-\frac{\lambda_0}{2}\right),
  \label{eq:Fdef}
\end{equation}
\begin{equation}
  K\equiv R_{\mu\nu}^2-\tfrac{1}{2}R^2,\qquad
  S\equiv 8R^{\mu\nu}R_{\mu\alpha}R^{\alpha}{}_{\nu}-6RR_{\mu\nu}^2+R^3,
  \label{eq:KSdef}
\end{equation}
and
\begin{equation}
  F_R\equiv\frac{\partial F}{\partial R}
  =-\frac{\sigma}{4m^2}\!\left[F+\left(1-\frac{\lambda_0}{2}\right)\right].
  \label{eq:FRdef}
\end{equation}
Varying~\eqref{eq:Faction} in the presence of a matter sector yields the field equations~\cite{cfunctions}
\begin{align}
  -\frac{\kappa^2}{8m^2}\,T_{\mu\nu}
  =\;&-\tfrac{1}{2}F g_{\mu\nu}
  +\left(g_{\mu\nu}\Box-\nabla_\mu\nabla_\nu\right)F_R
  +F_R R_{\mu\nu}
  \nonumber\\
  &-\frac{\sigma}{m^2}\Big\{
   2\nabla_\alpha\nabla_\mu\!\left(F_R R^{\alpha}{}_{\nu}\right)
   -g_{\mu\nu}\nabla_\beta\nabla_\alpha\!\left(F_R R^{\alpha\beta}\right)
   -\Box\!\left(F_R R_{\mu\nu}\right)
   -2F_R R^{\alpha}{}_{\nu}R_{\mu\alpha}
   \nonumber\\
  &\hphantom{-\frac{\sigma}{m^2}\Big\{}
   +g_{\mu\nu}\Box\!\left(F_R R\right)
   -\nabla_\mu\nabla_\nu\!\left(F_R R\right)
   +F_R R R_{\mu\nu}\Big\}
   \nonumber\\
  &-\frac{1}{2m^4}\Big\{
   4F_R R^{\rho}{}_{\mu}R_{\rho\alpha}R^{\alpha}{}_{\nu}
   +2g_{\mu\nu}\nabla_\alpha\nabla_\beta\!\left(F_R R^{\beta\rho}R^{\alpha}{}_{\rho}\right)
   +2\Box\!\left(F_R R^{\rho}{}_{\nu}R_{\mu\rho}\right)
   \nonumber\\
  &\hphantom{-\frac{1}{2m^4}\Big\{}
   -4\nabla_\alpha\nabla_\mu\!\left(F_R R^{\rho}{}_{\nu}R^{\alpha}{}_{\rho}\right)
   +2\nabla_\alpha\nabla_\mu\!\left(F_R R R^{\alpha}{}_{\nu}\right)
   -g_{\mu\nu}\nabla_\alpha\nabla_\beta\!\left(F_R R R^{\alpha\beta}\right)
   \nonumber\\
  &\hphantom{-\frac{1}{2m^4}\Big\{}
   -\Box\!\left(F_R R R_{\mu\nu}\right)
   -2F_R R R^{\rho}{}_{\nu}R_{\mu\rho}
   -g_{\mu\nu}\Box\!\left(F_R R_{\alpha\beta}^2\right)
   +\nabla_\nu\nabla_\mu\!\left(F_R R_{\alpha\beta}^2\right)
   \nonumber\\
  &\hphantom{-\frac{1}{2m^4}\Big\{}
   -F_R R_{\alpha\beta}^2 R_{\mu\nu}
   +\tfrac{1}{2}g_{\mu\nu}\Box\!\left(F_R R^2\right)
   -\tfrac{1}{2}\nabla_\mu\nabla_\nu\!\left(F_R R^2\right)
   +\tfrac{1}{2}F_R R^2 R_{\mu\nu}\Big\}.
  \label{eq:detEOM}
\end{align}
By contrast, the auxiliary-form equation~\eqref{eq:gEOM} below is much more compact;
the densitized auxiliary metric $\mathcal{P}^{\mu\nu}$ resums the entire
$F_R$-weighted tower above.

\subsection{Small-curvature expansion}

With $X^{\mu}{}_{\nu}=\tfrac{\sigma}{m^2}G^{\mu}{}_{\nu}$,
\begin{equation}
  \sqrt{-\det A}=\sqrt{-g}\,\sqrt{\det(\delta^{\mu}{}_{\nu}+X^{\mu}{}_{\nu})},
\end{equation}
and using $\sqrt{\det(1+X)}=1+\half\Tr X+\tfrac18(\Tr X)^2-\tfrac14\Tr(X^2)
+O(X^3)$ together with the three-dimensional identities
$\Tr G=-\tfrac12 R$ and $\Tr(G^2)=R_{\mu\nu}R^{\mu\nu}-\tfrac14R^2$,
\begin{equation}
  \sqrt{\det\!\Big(\delta^{\mu}{}_{\nu}+\tfrac{\sigma}{m^2}G^{\mu}{}_{\nu}\Big)}
  = 1 - \frac{\sigma}{4m^2}R
  - \frac{1}{4m^4}\Big(R_{\mu\nu}R^{\mu\nu}-\tfrac{3}{8}R^2\Big) + O(R^3).
  \label{eq:expansion}
\end{equation}
With the overall normalization of~\eqref{eq:BINMG} this reproduces NMG at
quadratic order. The Born--Infeld determinant is therefore not an arbitrary
resummation: its first nontrivial terms are exactly those of the massive
spin-two theory. See the original works for the higher-curvature expansion. 

\section{The auxiliary-metric formulation}
\label{sec:aux}

\subsection{Linearizing the square root}

On the regular branch, the determinant can be represented by an auxiliary metric.
We write
\begin{equation}
  I_{\rm aux}[g,q] = -\frac{4m^2}{\kappa^2}\int \dd^3x\,
  \Big[\half\sqrt{-q}\,\big(q^{\mu\nu}\Aux(g)-1\big) - \beta\sqrt{-g}\Big],
  \label{eq:Iaux}
\end{equation}
where $q_{\mu\nu}$ is an independent auxiliary metric, $q^{\mu\nu}$ its inverse,
and $\sqrt{-q}=\sqrt{-\det q_{\mu\nu}}$ and $A_{\mu \nu}$ was defined in  (\ref{eq:Adef}). The field $q_{\mu\nu}$ appears without
derivatives, so its equation of motion is algebraic. Varying~\eqref{eq:Iaux} with
respect to $q^{\mu\nu}$ gives
\begin{equation}
  \Aux - \half q_{\mu\nu}\big(q^{\alpha\beta}A_{\alpha\beta}-1\big)=0.
  \label{eq:qeom}
\end{equation}
Tracing with $q^{\mu\nu}$ in three dimensions yields $q^{\mu\nu}\Aux=3$, and
feeding this back into~\eqref{eq:qeom} gives
\begin{equation}
  q_{\mu\nu} = \Aux(g) = g_{\mu\nu}+\frac{\sigma}{m^2}G_{\mu\nu}(g).
  \label{eq:qeqA}
\end{equation}
Substituting~\eqref{eq:qeqA} back into~\eqref{eq:Iaux} returns the determinant
action~\eqref{eq:BINMG}, since $\half\sqrt{-q}(q^{\mu\nu}\Aux-1)
=\half\sqrt{-q}(3-1)=\sqrt{-q}=\sqrt{-\det A}$. Thus the auxiliary and determinant
formulations are equivalent on the regular branch.

\begin{remark}[Resummation]
The single algebraic relation~\eqref{eq:qeqA} encodes the entire infinite
curvature series of the determinant~\eqref{eq:BINMG}. The auxiliary metric
$q_{\mu\nu}$ is the curvature-deformed metric $\Aux$, and the higher-derivative
metric theory is rewritten as a second-order system in the pair $(g_{\mu\nu},
q_{\mu\nu})$.
\end{remark}

\subsection{The metric equation of motion}
\label{subsec:metric-eom-main}

The auxiliary equation obtained by varying $q^{\mu\nu}$ gives the algebraic
relation $q_{\mu\nu}=A_{\mu\nu}(g)$.  The remaining field equation is obtained by
varying the physical metric while keeping the auxiliary metric fixed.  This
point is important: in the auxiliary theory $g_{\mu\nu}$ and $q_{\mu\nu}$ are
independent fields before the auxiliary equation is imposed.

It is useful to introduce the densitized inverse auxiliary metric
\begin{equation}
  P^{\mu\nu}:=\sqrt{-q}\,q^{\mu\nu}.
  \label{eq:Pdef-main}
\end{equation}
This object is a contravariant tensor density.  To write covariant derivatives
without ambiguity we also introduce the ordinary tensor
\begin{equation}
  \Phi^{\mu\nu}:=\frac{P^{\mu\nu}}{\sqrt{-g}},
  \qquad
  \Phi:=g_{\mu\nu}\Phi^{\mu\nu}.
  \label{eq:Phidef-main}
\end{equation}
In three dimensions the density $P^{\mu\nu}$ also gives a polynomial form of the
action.  Indeed,
\begin{equation}
  \det(P^{\mu\nu})
  = (\sqrt{-q})^3\det(q^{\mu\nu})
  = \frac{(\sqrt{-q})^3}{q}
  = -\sqrt{-q},
  \label{eq:sqrtP-main}
\end{equation}
where the last equality uses $q<0$ for a Lorentzian auxiliary metric in three
dimensions.  Thus the auxiliary action may be written as
\begin{equation}
  I_{\rm aux}[g,P]
  = -\frac{4m^2}{\kappa^2}\int d^3x
  \left[
  \frac12 P^{\mu\nu}A_{\mu\nu}(g)
  +\frac12\det(P^{\mu\nu})
  -\beta\sqrt{-g}
  \right].
  \label{eq:IauxP-main}
\end{equation}
The Born--Infeld determinant of $A_{\mu\nu}(g)$ is absent off shell; it reappears
only after the auxiliary field is eliminated.  The derivative dependence is now
localized in the single term $P^{\mu\nu}G_{\mu\nu}(g)$.

The metric field equation, with the derivation given in Appendix~\ref{app:metric-variation}, takes the compact form
\begin{equation}
  \Phi^{\mu\nu}-\beta g^{\mu\nu}
  +\frac{\sigma}{m^2}K^{\mu\nu}[\Phi]=0 .
  \label{eq:gEOM}
\end{equation}
Here $K^{\mu\nu}[\Phi]$ is the adjoint of the linearized Einstein tensor acting on
$\Phi^{\mu\nu}$:
\begin{align}
K^{\mu\nu}[\Phi]
=&\;\frac12\Big(
\nabla_\rho\nabla^\mu\Phi^{\rho\nu}
+\nabla_\rho\nabla^\nu\Phi^{\rho\mu}
-\Box\Phi^{\mu\nu}
-g^{\mu\nu}\nabla_\rho\nabla_\sigma\Phi^{\rho\sigma}
\Big)
\nonumber\\
&-\frac12\nabla^\mu\nabla^\nu\Phi
+\frac12 g^{\mu\nu}\Box\Phi
+\frac12\Phi R^{\mu\nu}
-\frac12R\Phi^{\mu\nu} .
\label{eq:Koperator-main}
\end{align}
  Notice the
relative sign of the cosmological term in \eqref{eq:gEOM}; it follows from
varying $-\beta\sqrt{-g}$ with respect to the covariant metric.

Equation~\eqref{eq:gEOM} is the main advantage of the auxiliary formulation.
Before eliminating $q_{\mu\nu}$ it is second order in the independent fields
$(g_{\mu\nu},q_{\mu\nu})$.  After using
\begin{equation}
q_{\mu\nu}=A_{\mu\nu}(g)=g_{\mu\nu}+\frac{\sigma}{m^2}G_{\mu\nu}(g),
\label{eq:qA-repeat-main}
\end{equation}
the tensor $\Phi^{\mu\nu}$ contains two derivatives of the physical metric and
\eqref{eq:gEOM} becomes the fourth-order BINMG metric equation.  Equivalently, on
shell,
\begin{equation}
  \Phi^{\mu\nu}\longrightarrow
  \Psi^{\mu\nu}:=\frac{\sqrt{-\det A}}{\sqrt{-g}}
  (A^{-1})^{\mu\nu},
  \label{eq:Psi-main}
\end{equation}
where $(A^{-1})^{\mu\nu}$ denotes the matrix inverse of $A_{\mu\nu}$, not the
tensor obtained by raising both indices with $g^{\mu\nu}$.  Substituting
\eqref{eq:Psi-main} into \eqref{eq:gEOM} gives precisely the metric equation
obtained by varying the original determinant action.  This is shown explicitly in
Appendix~\ref{app:metric-variation}.

\section{The unique maximally symmetric vacuum}
\label{sec:vacuum}

For a maximally symmetric background, $\bar R_{\mu\nu}=2\Lambda\bar g_{\mu\nu}$,
$\bar G_{\mu\nu}=-\Lambda\bar g_{\mu\nu}$, so
\begin{equation}
  \bar A_{\mu\nu} = \Big(1-\frac{\sigma\Lambda}{m^2}\Big)\bar g_{\mu\nu}
  = a\,\bar g_{\mu\nu},
  \qquad a := 1 - \frac{\sigma\Lambda}{m^2},
\end{equation}
and $\bar q_{\mu\nu}=a\,\bar g_{\mu\nu}$. Here the two equations of motion play
complementary roles. The $q$-equation~\eqref{eq:qeom}, $q_{\mu\nu}=A_{\mu\nu}$,
uses $\bar q^{\alpha\beta}\bar A_{\alpha\beta}=3$ in three dimensions and reduces
to $a-\tfrac12 a(3-1)=0$, satisfied identically for every $a$: it fixes only the
form $\bar q_{\mu\nu}=a\,\bar g_{\mu\nu}$, not the scale. The scale is fixed by
the metric equation, which on the background (using
$\sqrt{-\bar q}\,\bar q^{\mu\nu}=\sqrt a\,\sqrt{-\bar g}\,\bar g^{\mu\nu}$) gives
\begin{equation}
  \beta = \sqrt a, \qquad\text{i.e.}\qquad
  a = \beta^2, \quad \Lambda = \frac{m^2}{\sigma}\big(1-\beta^2\big).
  \label{eq:vacuum}
\end{equation}
This single root is the unique maximally symmetric vacuum of BINMG, established
previously from the determinant field equations~\cite{BINMG,cfunctions} and
recovered here immediately.

The endpoint $\beta=0$ ($\lambda_0=2$, $a=0$) is not a sector to be excluded by
hand but a degenerate coupling at which the theory ceases to be a massive
gravity as mentioned before. The $\sqrt{-g}$ term then drops out of~\eqref{eq:Iaux}, the background
$\bar q_{\mu\nu}=\beta^2\bar g_{\mu\nu}$ vanishes, and $q^{\mu\nu}$, $\sqrt{-q}$
are undefined; equivalently, the Pauli--Fierz mass $M^2_{\rm BI}=-\sigma
m^2\beta^2$ (Section~\ref{sec:linear}) vanishes, so the graviton becomes massless
and propagates no local mode in three dimensions. Thus $\beta\neq0$ is the
condition for the action to define a massive gravity, not an extra assumption,
and we work throughout on $a=\beta^2>0$. The analogous tuning to a unique vacuum
with a controlled massless limit appears in higher dimensions
in~\cite{BIunique,BImassless4d}.

The contrast with NMG is sharp. The quadratic theory~\eqref{eq:NMG} has two
constant-curvature vacua $\Lambda_\pm=-2m^2(\sigma\pm\sqrt{1+\lambda_0})$, and the
attendant degeneracy is a known difficulty, since gravitational energy cannot be
compared across spaces asymptotic to different constant-curvature backgrounds.
The Born--Infeld resummation removes it: $\beta=\sqrt a$ selects one root, with
the degenerate partner absent.

\section{Linearized spectrum from the auxiliary action}
\label{sec:linear}

We now compute the linearized spectrum directly from~\eqref{eq:Iaux}, without
using the eliminated determinant form. We follow \cite{BHT2} and \cite{Tekin}. On the vacuum $\bar q_{\mu\nu}=a\bar
g_{\mu\nu}$ with $a=\beta^2>0$. Write $g_{\mu\nu}=\bar g_{\mu\nu}+\kappa
h_{\mu\nu}$ and expand the auxiliary metric about its own background, separating
the part that follows the metric perturbation from the genuinely auxiliary
fluctuation,
\begin{equation}
  q_{\mu\nu} = a\big(\bar g_{\mu\nu}+\kappa h_{\mu\nu}\big)
  + \kappa\,\chi_{\mu\nu} + O(\kappa^2).
\end{equation}
Under a linearized diffeomorphism $\chi_{\mu\nu}$ is gauge invariant. As in the
NMG auxiliary calculation it is convenient to trace-reverse it,
$f_{\mu\nu}:=\chi_{\mu\nu}-\half\bar g_{\mu\nu}\chi$.

Expanding the auxiliary density to second order and using the unique-vacuum
relation $\beta=\sqrt a$ to cancel the first-order terms, the quadratic action
separates into a metric sector and an auxiliary sector. After removing the mixing
by a shift $h_{\mu\nu}=\tilde h_{\mu\nu}+\tfrac{2}{a}f_{\mu\nu}$, one obtains
\begin{equation}
  \mathcal{L}^{(2)}_{\rm BINMG} = -\frac{\sigma\sqrt a}{2}\,
  \tilde h^{\mu\nu}\mathcal{G}_{\mu\nu}(\tilde h)
  + \mathcal{L}_f,
\end{equation}
with the auxiliary-sector Lagrangian
\begin{equation}
  \mathcal{L}_f = \frac{2\sigma}{a^{3/2}}\,f^{\mu\nu}\mathcal{G}_{\mu\nu}(f)
  - \frac{m^2}{\sqrt a}\big(f_{\mu\nu}f^{\mu\nu}-f^2\big),
  \label{eq:Lf}
\end{equation}
where $\mathcal{G}_{\mu\nu}$ is the cosmological linearized Einstein operator on
the maximally symmetric background \cite{DeserTekin}. The field $\tilde h_{\mu\nu}$ is governed by
three-dimensional linearized Einstein gravity and carries no local degrees of
freedom. The field $f_{\mu\nu}$ obeys a Pauli--Fierz action with mass
\begin{equation}
  M^2_{\rm BI} = -\sigma m^2 a = -\sigma m^2 + \Lambda,
  \label{eq:mass}
\end{equation}
and carries the two local helicities of a single massive spin-two excitation.

Two points deserve emphasis. First, the kinetic coefficient $2\sigma a^{-3/2}$ is
finite and nonzero for any $a=\beta^2>0$, and the mass~\eqref{eq:mass} is finite
and nonzero; the linearized theory at the vacuum is healthy, with no vanishing
normalization and no strong coupling. $\sigma = -1$ corresponds to the non-ghost sign. Second, the mass term in~\eqref{eq:Lf} has
the exact Fierz--Pauli structure $(f_{\mu\nu}f^{\mu\nu}-f^2)$ with relative
coefficient $-1$. In NMG this tuning is built into the action by hand
through~\eqref{eq:K}; here it is \emph{generated} by the Born--Infeld potential.
The auxiliary formulation thus shows that the ghost-freedom of the linearized
spectrum is an output of the determinant structure rather than an input. The
analysis refers throughout to $a=\beta^2>0$; as discussed in
Section~\ref{sec:vacuum}, the endpoint $a=0$ is a degenerate coupling at which the
theory is nondynamical, not a sector with a missing mode.

\section{Comparison: the NMG auxiliary field versus the BINMG auxiliary metric}
\label{sec:comparison}

It is worth setting the two auxiliary constructions side by side, since they
illustrate two different roles that a symmetric rank-two auxiliary variable can
play. At the level of tensor type, the distinction should not be overstated:
an auxiliary metric is a symmetric rank-two tensor restricted to a
nondegenerate sector of fixed signature. The relevant difference lies in the
way the auxiliary variable enters the action and in the purpose that it serves.

In NMG~\eqref{eq:NMGaux}, the auxiliary tensor \(f_{\mu\nu}\) is introduced on
top of the physical metric. Its indices are raised and lowered using
\(g_{\mu\nu}\), and it enters through a polynomial coupling to the Einstein
tensor together with an explicitly Fierz--Pauli-tuned quadratic potential.
Its purpose is to lower the derivative order of an already polynomial
curvature action.

In BINMG~\eqref{eq:Iaux}, the symmetric tensor \(q_{\mu\nu}\) is restricted to
the nondegenerate Lorentzian sector and is therefore naturally treated as an
auxiliary metric. It enters through its own inverse \(q^{\mu\nu}\) and volume
density \(\sqrt{-q}\). These structures are essential: they linearize the
Born--Infeld square root and resum the infinite curvature series into the
single algebraic relation~\eqref{eq:qeqA}. The Fierz--Pauli structure is not
inserted into the potential a priori but emerges in the quadratic expansion,
as shown in Section~\ref{sec:linear}.

The differences are summarized as follows.
\begin{itemize}[leftmargin=1.4em]
\item \emph{Auxiliary variable and its domain.} 
Both constructions use a symmetric rank-two auxiliary field. In NMG,
\(f_{\mu\nu}\) is used as an ordinary tensor whose contractions are formed with
the physical metric. In BINMG, \(q_{\mu\nu}\) is restricted to the
nondegenerate Lorentzian sector and enters through its own inverse and
determinant. It is in this precise sense that we refer to \(q_{\mu\nu}\) as an
auxiliary metric.
\item \emph{Potential.} NMG: a Fierz--Pauli mass term, quadratic and FP-tuned by
construction. BINMG: a Born--Infeld potential, which generates the FP tuning at
quadratic order as a derived consequence.
\item \emph{What is resummed.} NMG: nothing; the theory is already polynomial and
the auxiliary field truncates at finite order. BINMG: the entire infinite
curvature series, collapsed into~\eqref{eq:qeqA}.
\item \emph{Vacuum structure.} NMG: two degenerate vacua $\Lambda_\pm$. BINMG: a
unique vacuum $a=\beta^2$, with $a=0$ excluded; the auxiliary form makes the
uniqueness manifest.
\item \emph{Densitization.} The polynomial form via $P^{\mu\nu}=\sqrt{-q}\,
q^{\mu\nu}$ is specific to BINMG; NMG needs no such device.
\end{itemize}

The bimetric resemblance is not merely formal. The auxiliary action~\eqref{eq:Iaux}
is a two-metric system in which $q_{\mu\nu}$ carries no kinetic term and is fixed
by~\eqref{eq:qeqA}. This resemblance should be understood carefully.  The auxiliary metric in
BINMG has no independent Einstein--Hilbert kinetic term, so the theory is not a
generic bimetric gravity.  Nevertheless, the form of the action is naturally
compared with the scaling limits of ghost-free bigravity discussed in
Refs.~\cite{PaulosTolley,deRham3d}.  That comparison is suggestive and useful, but
it is not used here as a substitute for the auxiliary derivation.

\section{A basic application: charges and BTZ thermodynamics}

The auxiliary formulation is most useful when the auxiliary metric becomes
algebraically simple. This happens on locally anti-de Sitter configurations,
including the BTZ black hole. On such backgrounds
\begin{equation}
q_{\mu\nu} = A_{\mu\nu}(g) = a\, g_{\mu\nu}, \qquad
a = 1 - \frac{\sigma\Lambda}{m^2} = \beta^2,
\label{eq:q-AdS}
\end{equation}
and therefore
\begin{equation}
\sqrt{-q}\, q^{\mu\nu} = \sqrt{a}\, \sqrt{-g}\, g^{\mu\nu}.
\label{eq:Pmunu-AdS}
\end{equation}
The derivative part of the auxiliary action is
\begin{equation}
-\frac{2\sigma}{\kappa^2} \int d^3x\, \sqrt{-q}\, q^{\mu\nu} G_{\mu\nu}(g).
\label{eq:derivative-part}
\end{equation}
Using $g^{\mu\nu}G_{\mu\nu} = -R/2$ in three dimensions, this term reduces on
the locally AdS branch to the Einstein--Hilbert term with the effective
coupling
\begin{equation}
\frac{1}{\kappa^2_{\rm eff}} = \frac{\sigma\sqrt{a}}{\kappa^2}.
\label{eq:keff}
\end{equation}

 Once $q_{\mu\nu}$ is replaced by its on-shell
value \eqref{eq:q-AdS}, the curvature-dependent part of the action \emph{is}
the Einstein--Hilbert action, with no residual higher-curvature structure to
track --- the auxiliary metric has already absorbed it. Every quantity that
follows from the Einstein--Hilbert action with effective coupling
\eqref{eq:keff} is therefore available immediately, by substitution, rather
than by repeating a curvature expansion or a determinant variation on the
BTZ background.

For an asymptotic Killing vector $\bar\xi^\mu$ one obtains \cite{DeserTekin,Abbott,DeserTekin2}
\begin{equation}
Q_{\rm BINMG}[\bar\xi] = \sigma\sqrt{a}\, Q_{\rm Einstein}[\bar\xi],
\label{eq:charge-general}
\end{equation}
and hence for BTZ
\begin{equation}
M_{\rm BINMG} = \sigma\sqrt{a}\, M_{\rm BTZ}, \qquad
J_{\rm BINMG} = \sigma\sqrt{a}\, J_{\rm BTZ}.
\label{eq:MJ-BTZ}
\end{equation}
The same factor multiplies the Brown--Henneaux central charge \cite{Brown},
\begin{equation}
c_+ = c_- = \frac{3\ell}{2G}\, \sigma\sqrt{a},
\label{eq:central-charge}
\end{equation}
and the BTZ entropy,
\begin{equation}
S_{\rm BINMG} = \sigma\sqrt{a}\, \frac{2\pi r_+}{4G}.
\label{eq:BTZ-entropy}
\end{equation}

These results agree with the direct Wald-entropy computation for the
determinant form of the theory carried out in \cite{Ozen}, where the
same overall factor was obtained from the curvature derivatives of the
Lagrangian rather than from an auxiliary field. The agreement is a useful
check on the formulation: the algebraic shortcut
\eqref{eq:q-AdS}--\eqref{eq:keff} reproduces, in three lines, what otherwise
requires the Wald formula  \cite{Wald} and an explicit evaluation of
$\partial\mathcal{L}/\partial R_{\mu\nu}$ on the AdS background.

The same shortcut extends beyond what is computed here. Any local quantity
that is built covariantly from the metric and reduces, on a locally AdS
background, to a multiple of its Einstein-gravity counterpart, will pick up
the same factor $\sigma\sqrt{a}$ --- boundary stress tensors, holographic
Weyl anomalies, and one-point functions of the boundary CFT all fall into
this pattern, since they are computed from the same effective
Einstein--Hilbert sector \eqref{eq:derivative-part}--\eqref{eq:keff} rather
than from the original determinant action. A full boundary-term analysis
and holographic renormalization can also be developed in this language, but
those questions require a separate treatment and will not be pursued in the
present paper.

The sign of $\sigma\sqrt a$ is fixed by the bulk mass-sign condition of
Section~6; the resulting sign pattern for the boundary charges reflects the
same bulk--boundary unitarity tension familiar from TMG~\cite{Li}
and NMG~\cite{cfunctions}.

\section{Outlook: the nonlinear degree-of-freedom count}
\label{sec:nonlinear-outlook}

The auxiliary formulation was constructed with the nonlinear Hamiltonian
degree-of-freedom problem in view, and it is the natural setting for that
analysis. We indicate here precisely how it reduces the problem and what
remains, leaving the full treatment to a separate study \cite{BTekin}.

An earlier attempt to attack the BINMG Hamiltonian problem directly was made in \cite{Dogru}.  That thesis reviewed Dirac constraint analysis and applied the
ADM method to BINMG using a general \(f(R_{\mu\nu\rho\sigma})\) auxiliary-tensor
formalism.  It correctly identified the central obstruction: after many
constraints were classified, the nonlinear degree-of-freedom count depended on
the classification of a remaining pair of constraints.  The thesis did not
complete this classification; it explicitly left open the alternative between a
nonlinear two-degree-of-freedom theory and a theory with a third mode absent in
the linearized analysis. The problem turned out to be too difficult for such a direct attack.

In the polynomial form~\eqref{eq:IauxP-main}, the only derivative coupling is
$P^{\mu\nu}G_{\mu\nu}(g)$. After an ADM decomposition of the physical metric, the
densitized auxiliary metric splits into a normal-normal component, a mixed
normal-spatial component, and a spatial block. On the regular branch where the
spatial block is invertible, the mixed component is algebraic and is eliminated,
while the normal component remains linear in the Hamiltonian and thus imposes a
scalar constraint. Preservation of this scalar constraint generates a secondary
scalar constraint, and the would-be extra scalar mode (the Boulware-Deser ghost) of generic nonlinear
massive gravity~\cite{BoulwareDeser} is removed precisely when this scalar pair is
second class. The three first-class diffeomorphism constraints, together with such
a second-class scalar pair, leave two local degrees of freedom, matching the
linearized count of Section~\ref{sec:linear}. For the NMG case, a similar analysis, basically what we learned from \cite{Dirac,HT} was done in 
\cite{Blagovic, Sadegh}.

\section{Conclusions}

We have developed the auxiliary-metric formulation of Born–Infeld New Massive
Gravity.  The determinant action is replaced, on the regular Born–Infeld branch,
by an exactly equivalent action containing an independent auxiliary metric
$q_{\mu\nu}$.  Its equation of motion,
\begin{equation}
q_{\mu\nu}=g_{\mu\nu}+\frac{\sigma}{m^2}G_{\mu\nu}(g),
\end{equation}
recovers the determinant formulation after the auxiliary field is eliminated.
Before this elimination, however, the curvature-dependence is linear.  In the
densitized variable $P^{\mu\nu}=\sqrt{-q}\,q^{\mu\nu}$ the action is polynomial in
three dimensions, and the derivative content is carried by the single coupling
$P^{\mu\nu}G_{\mu\nu}(g)$.

This formulation gives a simple explanation of several known facts about BINMG.
The maximally symmetric vacuum is unique: on such a background, the auxiliary
metric is proportional to the physical metric, and the metric equation fixes the
proportionality factor to be $a=\beta^2$.  The linearized theory around this
vacuum contains one Pauli–Fierz massive spin–2 field.  In this derivation, the
Fierz–Pauli tuning is not inserted by hand; it is generated by the Born
–Infeld auxiliary potential.  On locally AdS configurations, the auxiliary metric is again
proportional to the physical metric, and the conserved charges, BTZ mass and
angular momentum, Brown–Henneaux central charge, and BTZ entropy reduce to the
corresponding Einstein quantities multiplied by $\sigma\sqrt a$, in agreement
with the direct Wald-entropy computation of~\cite{Ozen}; the same
algebraic substitution gives immediate access to any further locally AdS
observable of this type, such as boundary stress tensors and holographic Weyl
anomalies, without repeating the underlying curvature expansion.

The nonlinear degree-of-freedom count remains the important next problem.  The
auxiliary formulation does not by itself prove the absence of a nonlinear
Boulware–Deser--type scalar; what it does is isolate the problem in the right
variables.  The ADM decomposition of the polynomial action shows where the
candidate scalar constraint arises and which scalar constraint bracket must be
classified.  A complete Dirac analysis of that bracket, including possible
singular branches, will be given elsewhere.

The main lesson is that BINMG is not merely a compact determinant action.  It has
a hidden auxiliary-metric structure that makes the vacuum, the linearized
spectrum, and the rescaled AdS charges transparent, while providing a cleaner starting point for the genuinely nonlinear Hamiltonian analysis.

\appendix

\section{Metric variation of the auxiliary action}
\label{app:metric-variation}

In this appendix, we derive the metric field equation quoted in
Section~\ref{subsec:metric-eom-main}.  The derivation is included because the
compact equation in the main text is one of the useful outputs of the auxiliary
formulation, and because the signs are easiest to control if the auxiliary metric
is kept fixed until the end.

We start from
\begin{equation}
I_{\rm aux}[g,q]
=-\frac{4m^2}{\kappa^2}\int d^3x
\left\{
\frac12\sqrt{-q}\left[
q^{\mu\nu}\left(g_{\mu\nu}+\frac{\sigma}{m^2}G_{\mu\nu}(g)\right)-1
\right]
-\beta\sqrt{-g}
\right\}.
\label{eq:app-Iaux}
\end{equation}
The independent variables are $g_{\mu\nu}$ and $q_{\mu\nu}$.  In the metric
variation we therefore keep
\begin{equation}
\delta_g q_{\mu\nu}=0,
\qquad
\delta_g q^{\mu\nu}=0,
\qquad
\delta_g\sqrt{-q}=0.
\label{eq:app-fixed-q}
\end{equation}
Define the density
\begin{equation}
P^{\mu\nu}:=\sqrt{-q}\,q^{\mu\nu}
\label{eq:app-Pdef}
\end{equation}
and the corresponding ordinary tensor
\begin{equation}
\Phi^{\mu\nu}:=\frac{P^{\mu\nu}}{\sqrt{-g}}.
\label{eq:app-Phidef}
\end{equation}
Let
\begin{equation}
h_{\mu\nu}:=\delta g_{\mu\nu},
\qquad
h:=g^{\mu\nu}h_{\mu\nu}.
\label{eq:app-hdef}
\end{equation}
Then
\begin{equation}
\delta_g\sqrt{-g}=\frac12\sqrt{-g}\,g^{\mu\nu}h_{\mu\nu}.
\label{eq:app-dsqrtg}
\end{equation}
The variation of \eqref{eq:app-Iaux} is
\begin{align}
\delta_g I_{\rm aux}
=&-\frac{2m^2}{\kappa^2}\int d^3x\sqrt{-g}
\left(\Phi^{\mu\nu}-\beta g^{\mu\nu}\right)h_{\mu\nu}
\nonumber\\
&-\frac{2\sigma}{\kappa^2}\int d^3x\sqrt{-g}\,
\Phi^{\mu\nu}\delta G_{\mu\nu}.
\label{eq:app-var-start}
\end{align}
Thus the only nontrivial step is to integrate the last term by parts.

For a covariant metric variation one has
\begin{equation}
\delta g^{\mu\nu}=-h^{\mu\nu},
\qquad
h^{\mu\nu}:=g^{\mu\alpha}g^{\nu\beta}h_{\alpha\beta},
\label{eq:app-dginv}
\end{equation}
and
\begin{equation}
\delta\Gamma^\rho{}_{\mu\nu}
=\frac12 g^{\rho\lambda}
\left(\nabla_\mu h_{\nu\lambda}
+\nabla_\nu h_{\mu\lambda}
-\nabla_\lambda h_{\mu\nu}\right).
\label{eq:app-dGamma}
\end{equation}
Therefore
\begin{equation}
\delta R_{\mu\nu}
=\frac12\left(
\nabla_\rho\nabla_\mu h^\rho{}_{\nu}
+\nabla_\rho\nabla_\nu h^\rho{}_{\mu}
-\Box h_{\mu\nu}
-\nabla_\mu\nabla_\nu h
\right),
\label{eq:app-dRicci}
\end{equation}
and
\begin{equation}
\delta R
=-R^{\mu\nu}h_{\mu\nu}+\nabla_\mu\nabla_\nu h^{\mu\nu}-\Box h.
\label{eq:app-dR}
\end{equation}
Since
\begin{equation}
G_{\mu\nu}=R_{\mu\nu}-\frac12g_{\mu\nu}R,
\end{equation}
we have
\begin{equation}
\delta G_{\mu\nu}
=\delta R_{\mu\nu}-\frac12R h_{\mu\nu}-\frac12g_{\mu\nu}\delta R.
\label{eq:app-dG}
\end{equation}
After integrating by parts and dropping boundary terms, one obtains
\begin{equation}
\int d^3x\sqrt{-g}\,\Phi^{\mu\nu}\delta G_{\mu\nu}
=
\int d^3x\sqrt{-g}\,K^{\mu\nu}[\Phi]h_{\mu\nu},
\label{eq:app-Kidentity}
\end{equation}
where
\begin{align}
K^{\mu\nu}[\Phi]
=&\;\frac12\Big(
\nabla_\rho\nabla^\mu\Phi^{\rho\nu}
+\nabla_\rho\nabla^\nu\Phi^{\rho\mu}
-\Box\Phi^{\mu\nu}
-g^{\mu\nu}\nabla_\rho\nabla_\sigma\Phi^{\rho\sigma}
\Big)
\nonumber\\
&-\frac12\nabla^\mu\nabla^\nu\Phi
+\frac12g^{\mu\nu}\Box\Phi
+\frac12\Phi R^{\mu\nu}
-\frac12R\Phi^{\mu\nu} .
\label{eq:app-Koperator}
\end{align}
Substitution into \eqref{eq:app-var-start} gives
\begin{equation}
\delta_g I_{\rm aux}
=-\frac{2m^2}{\kappa^2}\int d^3x\sqrt{-g}
\left[
\Phi^{\mu\nu}-\beta g^{\mu\nu}
+\frac{\sigma}{m^2}K^{\mu\nu}[\Phi]
\right]h_{\mu\nu}.
\label{eq:app-final-var}
\end{equation}
Since $h_{\mu\nu}$ is arbitrary, the metric equation is
\begin{equation}
\Phi^{\mu\nu}-\beta g^{\mu\nu}
+\frac{\sigma}{m^2}K^{\mu\nu}[\Phi]=0.
\label{eq:app-metric-eq}
\end{equation}
This is Eq.~\eqref{eq:gEOM} of the main text.

Finally, imposing the auxiliary equation gives
\begin{equation}
q_{\mu\nu}=A_{\mu\nu}(g),
\qquad
\Phi^{\mu\nu}\rightarrow
\Psi^{\mu\nu}:=\frac{\sqrt{-\det A}}{\sqrt{-g}}(A^{-1})^{\mu\nu}.
\label{eq:app-Phi-to-Psi}
\end{equation}
The determinant action varies as
\begin{align}
\delta I_{\rm BINMG}
=&-\frac{4m^2}{\kappa^2}\int d^3x
\left[
\frac12\sqrt{-\det A}\,(A^{-1})^{\mu\nu}
\left(h_{\mu\nu}+\frac{\sigma}{m^2}\delta G_{\mu\nu}\right)
-\frac12\beta\sqrt{-g}\,g^{\mu\nu}h_{\mu\nu}
\right]
\nonumber\\
=&-\frac{2m^2}{\kappa^2}\int d^3x\sqrt{-g}
\left[
\Psi^{\mu\nu}-\beta g^{\mu\nu}
+\frac{\sigma}{m^2}K^{\mu\nu}[\Psi]
\right]h_{\mu\nu}.
\label{eq:app-det-var}
\end{align}
Thus, the auxiliary metric equation reproduces exactly the metric equation of the
determinant theory after $q_{\mu\nu}$ is eliminated.  The auxiliary formulation is
therefore equivalent not only at the level of the action but also at the level
of the field equations on the regular branch.

\section*{Acknowledgments} I thank my former students T.C. Sisman and I. Gullu, with whom I developed the BINMG theory and worked out many details, as well as finding generic $n$-dimensional Born-Infeld gravity theories with only a massless spin-2 excitation about their unique vacuum.  Some details can be found in Sisman's PhD thesis \cite{Tahsintez}. I also thank my former student M. Dogru, who carried out the difficult job of the Dirac constraint analysis for the BINMG in his master's thesis \cite{Dogru}; it was not finished; however, he almost got there.

\end{document}